\documentclass[12pt,preprint]{aastex}
\newcommand{\icrfext}{ICRF-Ext.1}
\newcommand{\catref}{http://www.nrao.edu/vlba/VCS1}

\slugcomment{Draft}

\shorttitle{The VLBA Calibrator Survey}
\shortauthors{Beasley et al.}

\begin{document}
\title{The VLBA Calibrator Survey - VCS1}

%% author stuff
\author{A.J. Beasley\altaffilmark{1}}
\affil{Owens Valley Radio Observatory,\\ California Institute of
Technology, P.O. Box 968, Big Pine CA 93513, USA}
\email{tbeasley@ovro.caltech.edu}
\author{D. Gordon}
\affil{Raytheon ITSS \& NASA Goddard Space Flight Center,\\
Code 926.9, Greenbelt MD 20771, USA}
\email{dgg@leo.gsfc.nasa.gov}
\author{A. B. Peck}
\affil{Harvard-Smithsonian Center for Astrophysics,\\
60 Garden St., Cambridge MA 02138, USA}
\email{apeck@cfa.harvard.edu}
\author{L. Petrov \& D. S. MacMillan}
\affil{NVI Inc. \& NASA/GSFC,
Code 926, Greenbelt MD 20771, USA}
\email{pet@leo.gsfc.nasa.gov, dsm@leo.gsfc.nasa.gov}
\author{E. B. Fomalont}
\affil{National Radio Astronomy Observatory,\\
520 Edgemont Rd., Charlottesville VA 22903, USA}
\email{efomalon@nrao.edu}
\and
\author{C. Ma}
\affil{NASA Goddard Space Flight Center,\\
Code 926, Greenbelt MD 20771, USA}
\email{cma@virgo.gsfc.nasa.gov}

%% alternative affiliations
\altaffiltext{1}{Combined Array for Research in Millimeter--wave Astronomy}

%% abstract
\begin{abstract}

A catalog containing milliarcsecond--accurate positions of 1332
extragalactic radio sources distributed over the northern sky is presented
-- the Very Long Baseline Array Calibrator Survey (VCS1). The positions
have been derived from astrometric analysis of dual--frequency 2.3 and 8.4 GHz
VLBA snapshot observations; in a majority of cases, images of the sources
are also available. These radio sources are suitable for use in geodetic
and astrometric experiments, and as phase--reference calibrators in
high--sensitivity astronomical imaging. The VCS1 is the largest
high--resolution radio survey ever undertaken, and triples the number of
sources available to the radio astronomy community for VLBI applications.
In addition to the astrometric role, this survey can be used in active
galactic nuclei, Galactic, gravitational lens and cosmological studies. The
VCS1 catalog is available at \catref.

\end{abstract}

%% Keywords

\section{Introduction}

Accurate celestial and terrestrial reference frames play an important role
in many areas of science.  Celestial reference frames have been used
historically for navigation (terrestrial and deep--space), time keeping
and for studying the dynamics of solar system, Galactic and extragalactic
objects. For over three decades, high--accuracy radio interferometry of
extragalactic radio sources using the techniques of Very Long Baseline
Interferometry (VLBI) has played a major role in defining these terrestrial
and celestial frames \citep{sovers98}.  Precise measurements of
orientation, rotation and deformation of the Earth's surface provides
unique information concerning its internal structure, and allows detailed
testing of geodynamic theories (e.g. plate tectonics) and studies of the
major geophysical fluids (the atmosphere, oceans and groundwater)
\citep{eubanks93}.

The International Earth Rotation Service (IERS -- http://www.iers.org) was
established in 1988 by the International Astronomical Union (IAU) and the
International Union of Geodesy and Geophysics (IUGG) to serve the
astronomical, geodetic and geophysical communities by providing an
International Celestial Reference System (ICRS) and an International
Terrestrial Reference System (ITRS) to define fundamental inertial frames
of reference for celestial and terrestrial positions. The current
realizations of the ICRS and ITRS are the International Celestial Reference
Frame (ICRF) \citep{ma98} and the International Terrestrial Reference Frame
(ITRF2000) \citep{altamimi02}. Physical connection of these two systems
depends on accurate measurement of two celestial angles (the offset in
longitude and obliquity of the celestial pole with respect to its position
defined by the conventional IAU precession/nutation models -- $\Delta\psi$
and $\Delta\epsilon$) and three Earth orientation parameters (EOPs): UT1
(changes in the length of day due to variations in the rotation of the
Earth), and the X \& Y polar motion offsets.  

The ICRF was adopted by the IAU as of January 1998 as the realization of
the ICRS, replacing the FK5 \citep{fricke88}. It is a kinematic reference
frame, based on the positions of distant quasars and active galactic
nuclei, rather than a dynamical reference frame such as the FK5 frame
(which incorporates the Earth's motion, the mean equator and the dynamical
equinox at some reference epoch). The origin of the ICRF is at the solar
system barycenter, and its axes are defined by the directions to a set of
extragalactic radio sources measured using VLBI. The ICRF axes are
consistent with the J2000 alignment of the FK5 to within the accuracy of
the FK5. The orientation of the ICRF is defined by the positions of the 212
defining sources included in the initial ICRF solution \citep{ma97}, with
an estimated accuracy of 0.25 milliarcseconds (mas). A further 396 sources
were included in \citet{ma98} as candidate ICRF sources and to improve the
catalog density, and subsequently 59 additional candidate sources were
presented in ICRF--Extension 1 (\icrfext) \citep{iers99}.  The data used to
define the ICRF included over 2.2 million VLBI observations made between
1979--1995. Significant improvement of the current ICRF will require VLBI
monitoring of the mas structure of the current sources, and the addition of
new sources to the frame solutions.

Astronomical imaging of weak radio sources using phase--referencing also
relies on a dense and accurate grid of bright radio sources. Phase
referencing \citep{beasley95} involves rapid ($\sim$minutes) antenna
position--switching between an astronomical target source and an adjacent
(1--5\arcdeg\, separation) calibrator. By interpolating antenna--based
phase, delay and rate corrections derived from the calibrator observations
to the weak target source, phase coherence can be extended indefinitely,
allowing longer integration times and thermal--noise--limited imaging. To
first order, unmodeled antenna--based geometric and electronic delays are
removed. Second--order errors in this process increase with (a) switching
time between observations of the calibrator, which depends on antenna slew
rates, source brightness and system sensitivity, and (b) the angular
distance to the calibrator (i.e. a breakdown of isoplanicity).  Dense
calibrator grids are required to minimize these errors.  Over the past five
years, phase--referencing has allowed imaging of weak radio sources
including GRBs \citep{taylor99}, radio stars \citep{beasley00} and
deep--field radio sources \citep{garrett01}.  Another advantage of phase
referencing is the astrometric registration of multi--epoch observations,
such as images of radio supernovae \citep{bartel00}, maser complexes
\citep{herrnstein99} and the Galactic center \citep{reid99}.

In this paper we present the results of a multi--epoch dual--frequency Very
Long Baseline Array (VLBA) survey -- the VLBA Calibrator Survey (VCS1) --
of 1811 VLBI sources identified on the basis of high--resolution Very Large
Array observations. Some 1332 of these sources have not been previously
measured in astrometric mode -- their positions are presented here as the
VCS1 catalog (see \catref). Future VLBA calibrator surveys (e.g. VCS2 --
\citet{fomalont01}) will build on this initial catalog. The goals of the
survey were: (a) to increase the surface density of known geodetic--grade
calibrators with mas--accurate positions in the northern sky,
providing candidate sources for future extensions of the ICRF; (b) to
facilitate routine phase--referencing to most regions of the northern and
equatorial sky, allowing high--resolution radio imaging of weak scientific
targets; and (c) to provide a uniform image database at 2.3 and 8.4 GHz for
use in scientific applications, including AGN \& gravitational lensing
studies, and cosmology. Milliarcsecond--accurate positions in the reference
frame of the ICRF and a selection of dual--frequency images of VCS1 sources
are presented. In Section 2 we describe the VCS1 sample and the survey
observations, and present a selection of survey images. In Sections 3 \& 4
we discuss the astrometric analysis of the VCS1 sample, and examine ongoing
efforts to use VCS1 for science and to expand it through the Galactic Plane
and to higher frequencies.

\section{Data \& Reduction}

\subsection{Observations}

The VCS1 observations were carried out in ten 24--hr sessions (epochs)
spanning the period August 1994 to August 1997.  A description of the VLBA
can be found in \citet{napier94}. VCS1 observations used the VLBA
dual--frequency geodetic mode, observing simultaneously at 2.3 \& 8.4 GHz.
This is enabled by a dichroic mirror permanently positioned over the 2.3
GHz receiver, reflecting the higher frequency radiation towards a
deployable reflector leading to the 8.4 GHz receiver. From each receiver
four baseband frequency channels (BBCs) were recorded over a large spanned
bandwidth (100 MHz at 2.3 GHz, 400 MHz at 8.4 GHz), to provide precise
measurements of group delays for astrometric processing. Relevant
parameters for the VCS1 epochs are indicated in Table~1. For epochs 1--4
the data recording rate was 64 Mb/s; growth of the VLBA operational
capabilities during 1996 allowed 128 Mb/s recording for epochs 6--10.
Typical VLBA antenna system temperatures and efficiencies at 2.3 \& 8.4 GHz
are $\sim$30--40\,K and 50\%. During each epoch, typically one or two
antennas exhibited poor performance due to local weather conditions.

To allow accurate imaging and position estimation, two or three
60--90\,second snapshot observations of each calibrator candidate were made.
The VCS1 sample was observed in declination strips, as indicated in
Table~1. In each declination strip the observations were sequenced to
maximize the number of VLBA antennas on source using customized scheduling
software.  During each epoch, six or seven periods of 30\,mins were used to
observe $\sim$10 geodetic--grade reference sources spanning a wide range of
elevations and azimuths to assist in the astrometric processing of the
survey. To schedule these reference sources, an elevation--weighted
simulated annealing algorithm was developed. For a given date and time,
this algorithm identifies sources visible to all VLBA antennas, generates a
preferred sample spanning a wide range of antenna elevations and azimuths,
giving higher weight to the outlying VLBA stations, i.e.  those involving
longer baselines which provide the highest astrometric accuracy. After
selection, the list of reference sources is ordered in time using a
simulated annealing algorithm \citep{press92} to give the minimum antenna
slew path through the sources, ignoring azimuth cable unwrapping.

\subsection{VCS1 Sample Selection}

A sample of candidate calibrators was selected from the Jodrell Bank -- VLA
Astrometric Survey (JVAS) \citep{patnaik92,browne98,wilkinson98,wrobel02},
an astrometric snapshot survey of compact radio sources performed with the
NRAO Very Large Array during the period 1990--1993. The primary goal of
JVAS was to provide phase calibrators with accurate positions for use in
radio astronomy, in particular for the MERLIN array, although the survey
has also been used to identify gravitationally--lensed systems, e.g.
\citet{king99}. The JVAS sample was originally derived from the NRAO Green
Bank 1.4 \& 5 GHz surveys \citep{condon85,condon86,condon89}, selecting
sources with 5 GHz flux densities greater than 200 milliJansky (mJy),
spectral indices $\alpha$ greater than -0.5 (S\,$\propto\nu^\alpha$), and
Galactic latitude $|b|$ greater than 2.5\arcdeg. These criteria were
adopted to select bright, compact extragalactic radio sources with
mas--scale core emission. The JVAS catalog contains 2118 sources in the
northern hemisphere; a southern hemisphere extension of $\sim$1000 sources
(0 -- -30\arcdeg) is in preparation \citep{wrobel02}.  The positional
accuracy of JVAS sources varies from 12 to 55 mas (1$\sigma$ errors),
depending on declination and observing conditions. The celestial frame in
which the JVAS positions have been specified is not formally defined, but
it appears consistent with the ICRF at the mas level. Systematic offsets
due to frame alignment and/or definition are considerably smaller than the
observational errors in the individual JVAS positions.

To identify a sample of calibrator candidates suitable for high--resolution
applications, we selected sources from the JVAS satisfying two criteria:
(a) JVAS structure estimates indicating point--like emission, suggesting
compact emission on scales of 10--30 mas; and (b) 8.4 flux density $\geq$
200 mJy (150 mJy in epochs 1 \& 2), i.e. suitably bright for
phase--referencing. An input sample of 1811 JVAS sources was selected
and observed during the VCS1 epochs. The positional accuracy of the JVAS
catalog is sufficient to provide initial correlation positions for VLBI
purposes. We note that the VCS1 sample is a representative but not
statistically--complete sample of compact flat--spectrum sources, due to
incompleteness in the original JVAS survey \citep{patnaik92} and possible
source variability since the Green Bank and JVAS surveys.

Throughout the survey, a sample of 57 bright geodetic--grade reference
sources were observed to provide estimates of epoch--to--epoch and absolute
positional accuracy, and to bootstrap atmospheric and instrumental error
estimates for the individual epochs. All of these sources are included in
the ICRF and \icrfext\, catalogs \citep{ma97}. A total of 387 (out of 667
cataloged) \icrfext\, sources were observed during this survey, and new
position estimates were made using dual--frequency VCS1 observations for 72
\icrfext\, sources with cataloged position errors greater than 0.5 mas rms.
Fig.~1 shows the celestial distribution of the 667 \icrfext\, sources (top)
and the 1332 VCS1 calibrators (bottom).

\subsection{Calibration \& Imaging}

Amplitude and initial phase calibration of the survey was done using the
National Radio Astronomy Observatory {\it Astronomical Imaging Processing
System} (AIPS) software. Absolute amplitude calibration is achieved 
using internal noise calibration sources at the VLBA antennas.
After fringe fitting, the amplitude and phase--calibrated data were exported
to FITS format for imaging purposes.  Automated imaging of the VCS1 sources
to enable on--line access was carried out using the Caltech {\it Difmap}
package. Starting with a point source model, iterations of {\it Clean} and
self--calibration procedures were carried out until noise--like residuals
were encountered, at which point a final naturally--weighted image was
produced.  A selection of the VCS1 candidate images is shown in Fig.~2. The
typical image rms is 2--3\,mJy, with a dynamic range of $\sim$30:1 or
better. Of the total 1811 sources observed, approximately 1300
sources (70\%) imaged automatically at 8.4 GHz.  Additional sources were
manually imaged.  Calibrator candidates which failed to image generally had
too little data, or were strongly resolved on a majority of VLBA baselines,
indicating source sizes of greater than 10--20 mas. The images and {\it
u--v} radius plots for the VCS1 can be accessed online via a graphical
search engine \citep{nraoonline}.

\section{Astrometry}

Astrometric processing of the VCS1 was performed at the NASA Goddard
Space Flight Center using {\it AIPS} and the {\it Calc/Solve} software
packages. Each epoch and each frequency band (2.3 and 8.4 GHz) was
processed independently in {\it AIPS}. Electronic phase offsets were applied
to the amplitude calibrated visibilities using either measured
phase tone phases (epochs 4--10) or using phase offsets determined
by fringe fitting a reference scan on a bright source (epochs 1--3).
Phase and amplitude variations across the individual BBCs were estimated
and corrected using a reference scan on a bright source.  Individual
baselines were then fringe fitted in {\it AIPS} to obtain residual
single-band delays, multi--band (bandwidth synthesis) delays, phase delay
rates, and fringe phases. These quantities were then combined with the
correlator model to obtain total group delays, delay rates, and phases.
Finally, the totals were shifted from geocentric quantities to the standard
geodetic/astrometric reference--station baseline observables, and were
written out in a form suitable for import into the {\it Calc/Solve}
analysis package. {\it Calc/Solve} is a geodetic/astrometric VLBI analysis
package, developed and maintained by the GSFC VLBI group and other partners
for the international geodetic/astrometric community.

{\it Calc} computes theoretical delays and delay rates using astronomical
and geophysical models following the IERS 1996 Conventions
\citep{mccarthy96}, and is used by numerous correlators around the world,
including the VLBA, Mark III, Mark IV, JIVE, DRAO and ATNF correlators.
{\it Solve} performs a least squares fit between the observed and
theoretical delays and delay rates, allowing numerous parameters to be
adjusted. Ionosphere--corrected linear combinations of the observed 2.3 and
8.4 GHz band group delays were used in the analysis.  Initially, 
{\it Solve} analyses were made of each epoch to resolve group delay
ambiguities, to eliminate outliers (2--4\% of observations), and to
estimate preliminary VCS1 source positions.  When all epochs were ready, a
final {\it Solve} analysis was made of all VCS1 epochs simultaneously to
estimate: (a) final source positions; (b) positions of all stations except
one assumed as a reference; (c) rate of change of UT1; (d) nutation
offsets; (e) troposphere path delays and station clock behavior; and (f)
east--west and north--south troposphere gradients and their derivatives.
Clocks and tropospheres were modeled by linear splines with 60\,min
intervals, with constraints on their rate of change imposed with reciprocal
weights of 50\,ps/hr and {$5 {\rm x} 10^{-14}$}\,s/s, respectively.

In total, 1811 radio sources were observed in the VCS1 campaign.  Two known
gravitational lenses, 0218+357 and 1830-211, as well as the radio source
0001-121 were excluded from the astrometric solution since the fringing
process for different observations apparently chose peaks which
corresponded to different source components.  To align the VCS1 positions
with the reference frame defined by \icrfext, the positions of 315
\icrfext\, reference sources observed were fixed at their catalog values
during the astrometric processing. These fixed sources have position error
ellipses with semi--major axes of 1.0 mas or less \citep{iers99}.

The positions of 1370 sources with three or more group delays at both 2.3
\& 8.4 GHz were estimated from the VCS1 data. An additional 34 sources had
fewer than two successful observations at 2.3 GHz but more than two good
observations at 8.4 GHz. A supplementary solution using only 8.4 GHz group
delays was made including these 34 sources. For 9 sources with fewer than
ten dual--frequency delays but considerably more 8.4 GHz delays, the
positions from the 8.4 GHz--only solution were also taken as being more
reliable.  Since VCS1 observations were made during the years of solar
minimum, errors due to uncorrected ionospheric delays in the 8.4 GHz data
were relatively small. The final VCS1 catalog contains positions of 1289
sources from the dual--frequency solutions, and 43 source positions derived
from the 8.4 GHz--only solution. 

In Table~2 the survey information for the first 10 sources in the VCS1
catalog is shown, including J2000 and IVS B1950 source names, J2000
positions and position errors, the correlation $r$ between the position
error estimates in right ascension and declination, the number of
observations included (N$_{\rm obs}$) and the solution origin (X/S --
solution incorporates both 2.3 \& 8.4 GHz group delays; X-only -- solution
incorporates only 8.4 GHz group delays; X+ -- solution incorporates only
8.4 GHz group delays: 2.3 GHz group delays were available, but the number
of 2.3 GHz group delays was significantly less than the number of 8.4 GHz
group delays). The parameter $r$ is a measure of the covariances of the
right ascension and declination position estimates, with high values of
$|r|$ ($\geq 0.5$) generally indicating that the array geometry was not
optimal during the observations of this source, leading to poor position
estimation.  There were 126 sources not detected at either band, or
yielding only one measurement of group delay, and were therefore not used
in the astrometric solution (these sources are listed in Table~3).  

The VCS1 catalog, and a catalog containing improved positions for the 72
\icrfext\, sources solved for in the VCS1 dual--frequency astrometric
analysis (Table~4), are available online at \catref. Complete versions of
Table~2 and Table~4 are also available in the electronic--only version of
the journal. 

\subsection{Error Analysis}

The observing methodology of the VCS1 was significantly different from that
used by traditional geodetic/astrometric VLBI experiments. The major source
of errors in geodetic/astrometric experiments is from fluctuations in the
troposphere path delays, which are estimated from the observations
themselves. In order to separate the troposphere from other parameters in
the least--squares solution, the VCS1 observing schedules included
observations of sources at both low and high elevation for each station
every hour, and to cover the sky as uniformly as possible. Fig.~3
presents the sky coverage for a VLBI antenna (the North Liberty VLBA
station) during a typical geodetic/astrometric experiment and during a VCS1
epoch.  The VCS1 was observed in declinations strips, leading to
significantly less uniform sky coverage.  Therefore, one could expect that
the VCS1 catalog positions may have some systematic errors due to poorer
estimates of troposphere zenith path delay and horizontal troposphere
gradients.

In order to investigate the presence of systematic errors of the VCS1
catalog a trial solution was made. The positions of all sources except the
57 reference sources were estimated, including 296 sources that were in
both the VCS1 and \icrfext\, source lists. Vectors of the differences in
positions between the trial solution and \icrfext\, for 272 sources with
semi--major error axes less than 2 mas are shown in Fig.~4.  No pattern is
apparent, suggesting that systematic effects in the VCS1 astrometric
analysis are small.

It has been found from analysis of different geodetic results
\citep{ryan93}, such as baseline lengths, Earth orientation parameters,
source positions, etc., that the formal errors of these estimated
parameters should be scaled up by the factor of 1.5.  Analysis of source
positions have indicated that the actual errors generally have a noise
floor \citep{ma98}. Therefore source positions errors in the ICRF
catalog were inflated:

\begin{equation}
\sigma_{r} = \sqrt{ (1.5 \cdot \sigma_{f})^2 + a^2 }
\label{errors_eqn}
\end{equation}

where $\sigma_{r}$ stands for the reported uncertainties, $\sigma_{f}$
stands for the formal errors and $a$ stands for the additive noise term.

For the VCS1 we have adopted a similar error model. We used the
differences between the trial solution and the \icrfext\, catalog and
obtained an additive noise term of 0.4 mas for VCS1 source positions
derived using ionosphere--free joint dual--frequency group delays, and a
noise term of 1.0 mas when 8.4 GHz--only delays were used. This additive
term makes $\chi^2$ per degree of freedom of the differences in source
positions between the trial solution and \icrfext\, approximately unity.
Reported uncertainties of source positions in the VCS1 catalog were
inflated using equation (\ref{errors_eqn}) and these values of the additive
noise terms.

A histogram of the distribution of the semi--major axes of the inflated
error ellipses is presented in Fig.~5. In the VCS1 catalog 53\% of sources
catalog have errors less than 1 mas and only 8\% of sources have errors
greater than 10 mas.  The median value of the VCS1 position error
distribution is $\sim$0.9 mas while the median error of the \icrfext\,
catalog is 0.5 mas. Typical geodetic/astrometric experiments focus on
source samples with simple or known brightness distribution, and have much
better sky coverage. The complex effects of undetermined and variable
source structure on astrometric positions have been studied recently
\citep{charlot90,fey00} and may be assessed for the VCS1 in the future
using the results of automated imaging and follow--up observations.
Multi--year phase-referencing observations using compact extragalactic
reference sources taken from VCS1, \icrfext1\, or other catalogs may be
affected by source evolution and variability at the sub-mas level, and
therefore careful monitoring of reference source structure or the use of
multiple reference sources is required in high-precision astrometric
applications.

\section{Discussion}

The mas--accurate positions for compact bright extragalactic radio sources
presented in this survey will enable phase--referencing VLBI imaging of weak
astronomical targets over large areas of the northern sky.  The combined
\icrfext\, and VCS1 sky density is sufficient to provide a calibrator
within 3\arcdeg\, of a random location north of -30\arcdeg\, declination
approximately 75\% of the time, and within 5\arcdeg\, in 96\% of cases.
Experiments requiring high astrometric accuracy, or in cases where no
suitable cataloged VLBI calibrator can be found within a few degrees, may
use weaker continuum sources close to the astronomical target if sufficient
data recording bandwidth is available. These weaker continuum sources may
be identified from lower--resolution surveys such as the NRAO VLA Sky Survey
\citep{condon98}.  Short and long--term source variability may vary the flux
densities of the VCS1 sources by tens of percent in extreme cases, so
weaker calibrators should be examined before use in critical applications.  

Use of the VCS1 in scientific studies has already commenced. The parent
survey to VCS1 -- JVAS -- has been extensively used to search for
gravitationally--lensed systems \citep{patnaik92b,king99}. The existence or
absence of gravitational--lens pairs on mas--scales can be used to place
limits on the cosmological abundance of supermassive compact objects in the
mass range $\sim10^6$\, to $10^8$\,M$_\sun$\, \citep{wilkinson01}. Current
studies based on samples of $\sim$300 sources indicate that such objects
cannot make up more than ~1\% of the closure density of the universe. This
suggests that a population of supermassive black holes forming soon after
the Big Bang does not contribute significantly to the dark matter content
of the Universe.

Examination of the VCS1 images to identify compact--symmetric objects
(CSOs) has been carried out \citep{peck00a}, doubling the number of these
objects available for follow--up observations, including the detection of
neutral hydrogen absorption towards the radio components of some CSOs
\citep{peck00b}, which may indicate the presence of an obscuring atomic
torus in the nucleus of these objects. CSOs have also been found to be
remarkably stable flux calibrators \citep{fassnacht01}, so larger samples
will prove valuable for VLA and VLBA flux monitoring experiments, such as
measuring time delays in gravitationally lensed sources.

Additional surveys based on the VCS1 to identify sources suitable as
geodetic and phase--referencing calibrators at high frequencies (22--90
GHz), and throughout the Galactic Plane, have recently commenced
\citep{fomalont01}.  Proper--motion studies of Galactic objects such as
pulsars \citep{brisken00} depend on dense grids of suitable low--frequency
calibrators at low Galactic declinations, which VCS1 does not cover.  The
results of these efforts should further benefit the geodetic and
astrometric communities.

At the present time, radio observations are the most accurate way to define
the ICRS; however in future optical observations will likely play an
important role. The Hipparcos stellar reference frame \citep{perryman97}
has been aligned with the ICRF to within 0.6 mas offset and  0.25 mas in
rotation at epoch 1991.25 \citep{kovalevsky97}, and represents the optical
realization of the ICRS. Over the next two decades, new optical
interferometers and astrometry missions such as the NASA Space
Interferometry Mission \citep{shao98}, the US Naval Observatory Full--Sky
Astrometric Mapping Explorer \citep{horner00} and the European Space
Agency's Global Astrometric Interferometer for Astrophysics
\citep{perryman01} will achieve microarcsecond positional accuracies,
requiring new definitions of the ICRS. 

\acknowledgments

The authors thank the many NRAO, GSFC and USNO staff members who offered
advice and assistance during this work. The National Radio Astronomy
Observatory is a facility of the National Science Foundation operated under
cooperative agreement by Associated Universities, Inc.. AP is grateful to
the New Mexico Space Grant Consortium for partial support during this
project. 

\clearpage

%\appendix
%\clearpage

%myrefs

\clearpage

\begin{deluxetable}{ccccccc}
\tabletypesize{\scriptsize}
\tablecaption{VCS1 Observations\label{tbl-1}}
\tablewidth{0pt}
\tablehead{
\colhead{Epoch} & \colhead{Date}   & \colhead{$\delta$\tablenotemark{a}}   & \colhead{N$_{\rm srcs}$\tablenotemark{b}} &
\colhead{Mb/s} & \colhead{Notes} }
\startdata
1  & 08/12/94 & 50\arcdeg -- 61\arcdeg & 216 &  64 & \\
2  & 05/19/95 & 61\arcdeg -- 79\arcdeg & 215 &  64 & \\
3  & 07/15/95 & 00\arcdeg -- 12\arcdeg & 216 &  64 & \\
4  & 03/13/96 & 12\arcdeg -- 24\arcdeg & 216 &  64 & (d) \\
5  & 05/15/96 & 24\arcdeg -- 34\arcdeg & 192 &  128 &  \\
6  & 06/07/96 & 34\arcdeg -- 44\arcdeg & 192 &  128 &  \\
7  & 08/10/96 & 44\arcdeg -- 51\arcdeg & 192 &  128 &  (d) \\
8  & 05/07/97 & 00\arcdeg -- -14\arcdeg & 192 &  128 &  \\
9  & 07/02/97 & -14\arcdeg -- -28\arcdeg & 191 &  128 &  \\
10 & 08/27/97\tablenotemark{c} & -18\arcdeg -- -30\arcdeg & 160 &  128 &  \\
   &    "      & 78\arcdeg -- 90\arcdeg & 32  & 128  &  \\
\enddata
\tablenotetext{a}{Declination range observed during each epoch.}
\tablenotetext{b}{Number of VCS1 candidates sources observed per
epoch. Approximately 10\% of the sample was observed in multiple epochs
to assess astrometric accuracy.}
\tablenotetext{c}{Two declination ranges were observed during this epoch.}
\tablenotetext{d} {Three epochs (01/02/96, 03/21/97 \& 04/04/97) were
abandoned due to poor weather at a majority of the VLBA sites.}
\end{deluxetable}

\clearpage

\begin{deluxetable}{ccccccccrrrcc}
\tabletypesize{\scriptsize}
\tablecaption{VCS1 Catalog Sample}
\tablewidth{0pt}
\tablehead{
\colhead{J2000 name} & \colhead{IVS B1950 name}   & 
\multicolumn{3}{c}{Right Ascension (J2000)} &  
\multicolumn{3}{c}{Declination (J2000)} & 
\colhead{$\sigma_{ra}$} & 
\colhead{$\sigma_{dec}$} & 
\colhead{$r$} & 
\colhead{N$_{\rm obs}$} & 
\colhead{Code}  \\ 
%\cline{3-5} \cline{6-8} \\
%
 & & 
\colhead{h} & \colhead{m} & \colhead{s}  & 
\colhead{$^\circ$} & \colhead{'} & \colhead{"} & 
\colhead{mas} & 
\colhead{mas} & 
 & & \\
}
\startdata
J0000$+$4054 & 2358$+$406 &  00 & 00 & 53.081551 & $+$40 & 54 & 01.79335 & 2.38 &   2.11 & $-0$.16  &     22 & X/S \\
J0001$-$1551 & 2358$-$161 &  00 & 01 & 05.328752 & $-$15 & 51 & 07.07583 & 0.62 &   0.95 & $-0$.75  &     58 & X/S \\
J0001$+$1914 & 2358$+$189 &  00 & 01 & 08.621564 & $+$19 & 14 & 33.80177 & 0.44 &   0.48 & $0$.00  &    101 & X/S \\
J0003$-$1927 & 0000$-$197 &  00 & 03 & 18.675008 & $-$19 & 27 & 22.35478 & 0.65 &   1.00 & $-0$.22  &     76 & X/S \\
J0003$+$2129 & 0000$+$212 &  00 & 03 & 19.350020 & $+$21 & 29 & 44.50760 & 0.74 &   1.21 & $-0$.26  &     35 & X/S \\
J0004$-$1148 & 0001$-$120 &  00 & 04 & 04.914996 & $-$11 & 48 & 58.38567 & 0.42 &   0.51 & $0$.05  &     81 & X/S \\
J0004$+$4615 & 0001$+$459 &  00 & 04 & 16.127657 & $+$46 & 15 & 17.96994 & 0.71 &   0.72 & $0$.10  &     75 & X/S \\
J0005$+$5428 & 0002$+$541 &  00 & 05 & 04.363490 & $+$54 & 28 & 24.92651 & 1.43 &   1.17 & $0$.46  &     60 & X/S \\
J0005$-$1648 & 0002$-$170 &  00 & 05 & 17.933794 & $-$16 & 48 & 04.67886 & 0.57 &   0.94 & $-0$.55  &     64 & X/S \\
J0005$+$0524 & 0002$+$051 &  00 & 05 & 20.215569 & $+$05 & 24 & 10.80084 & 2.10 &   2.22 & $-0$.10  &     26 & X/S \\
\enddata

\end{deluxetable}

\clearpage

\begin{deluxetable}{cccccc}
\tabletypesize{\scriptsize}
\tablehead{\multicolumn{6}{c}{JVAS J2000 source names}}
\tablecaption{VCS1 Observations\label{tbl-2}}
\tablecaption{VCS1 Weak or Non--detections}
\tablecolumns{6}
\tablewidth{0pc}
\startdata
J0003+4807 &  J0003-1149 & J0030+5904 & J0032+1953 & J0045+4555 & J0121+1127 \\
J0127+7323 &  J0145+5810 & J0151+5454 & J0214+5144 & J0221+3556 & J0308+6955 \\
J0332+6753 &  J0344+6518 & J0354+6621 & J0401+0413 & J0418+5457 & J0426+6825 \\
J0429+6710 &  J0448+5921 & J0505+6406 & J0524+7034 & J0544+5258 & J0610+7801 \\
J0612+6225 &  J0642+5247 & J0647+5446 & J0657+5741 & J0714+7408 & J0718+6651 \\
J0733+5605 &  J0749+5750 & J0752+5808 & J0754+7140 & J0754+5324 & J0757+6110 \\
J0809+5341 &  J0814+6431 & J0836+0052 & J0853+6828 & J0853+6722 & J0855+5751 \\
J0856+7146 &  J0907+6644 & J0917+6530 & J0921+1350 & J0929+7304 & J0943+6150 \\
J0948+0022 &  J1011+6529 & J1027+7428 & J1032+5610 & J1034+6832 & J1124+6555 \\
J1132+0034 &  J1148+5254 & J1154+5934 & J1156+7306 & J1200+5300 & J1203+6031 \\
J1204+5228 &  J1210+6422 & J1219+6344 & J1233+5026 & J1241+5458 & J1247+7124 \\
J1256+5652 &  J1258+5421 & J1316+6726 & J1330+5202 & J1340+6923 & J1350+6132 \\
J1353+6324 &  J1401+5835 & J1407+7628 & J1411+5917 & J1410+6216 & J1420+1703 \\
J1429+6316 &  J1448+5326 & J1451+6357 & J1507+5857 & J1520+5635 & J1524+7336 \\
J1542-0927 &  J1541+5348 & J1548+7845 & J1557+4522 & J1556+7420 & J1558+5625 \\
J1603+6945 &  J1610+7809 & J1629+6757 & J1628+7706 & J1651+5805 & J1656+5321 \\
J1705+7756 &  J1746+6421 & J1745+6703 & J1757+7539 & J1803+0934 & J1825+5753 \\
J1833-2103 &  J1850+4959 & J1928+6814 & J1926+7706 & J1930+5948 & J1938+6307 \\
J1952+4958 &  J2007+7452 & J2015+4628 & J2014+6553 & J2052+6858 & J2100+5612 \\
J2106+6004 &  J2129+6819 & J2203+7151 & J2209-2331 & J2223+6249 & J2232+6249 \\
J2307+1450 &  J2322+6911 & J2331+4522 & J2343+7003 & J2347+5142 & J2349+7517 \\
\enddata
\end{deluxetable}

\end{document}